# History-dependent nano-photoisomerization by optical near-field in photochromic single crystals

Yuji Arakawa[1], Kazuharu Uchiyama[1,*], Kingo Uchida[2], Makoto Naruse[3], and Hirokazu Hori[1]

[1] *University of Yamanashi, 4-3-11 Takeda, Kofu, Yamanashi 400-8511, Japan*

[2] *Ryukoku University, 1-5 Yokotani, Oe-cho, Seta, Otsu, Shiga 520-2194, Japan*

[3] *Department of Information Physics and Computing, Graduate School of Information Science and Technology, The University of Tokyo, 7-3-1 Bunkyo-ku, Tokyo 113-8656, Japan*

\* Corresponding author: kuchiyama@yamanashi.ac.jp

**Abstract**

*We demonstrate history-dependent or dynamic nano-photoisomerization by sequential formation of multiple memory pathways in photochromic crystals via optical near-field interactions. We observed the incident photons passing through the photoisomerization memory pathways by a double-probe optical near-field microscope, with one probe located on the front surface for local excitation and the other on the rear surface for near-field observation. By carrying out localized near-field excitation twice but at spatially different positions, we observed negatively correlated near-field output patterns between the first memory pathway and the second memory pathway. That is, the added memory pathway was formed exclusively to the previously formed memory pathway. We also confirmed that the first memory pathway was preserved after the second memory pathway was formed. This result indicates that photoisomerization by an optical near-field in diarylethene crystals has a history dependence, leading to brain-like dynamic information memorization using light–matter interactions on the nanometer-scale.*

**Introduction**

In recent years, computing based on von Neumann architectures has faced fundamental difficulties in solving complex problems [1]. Physical and spontaneous processes have the potential to be utilized for future intelligent devices and systems beyond conventional architectures [2,3]. Nowadays, such emerging system architectures are being intensively studied, ranging from neuromorphic computing [4,5] and nature-inspired annealing [6,7], to photonic decision making [8]. Generally speaking, we consider that these principles utilizing physical processes for computing fall under the concept of natural intelligence, which relates or associates complex phenomena or structures in natural systems with problems under study. In the literature, natural computations with amoebas [9] and decision making with single photons [10], for example, have been demonstrated for solving complex problems, using parallelism and history-dependent selection of alternatives by exploiting the non-triviality of physical systems. These studies indicate that the parallelism and history-dependence of the physical systems are promising resources for solving complex problems.

Functional computation in a physical system needs spontaneous memory structures embedded in the physical system. Meanwhile, persistent storage of information in the form of light has been known to be extremely difficult technologically [11,12], and overcoming the diffraction limit of light for high-density storage is also difficult [13]. Therefore, we focus on the crystal structure of photochromic molecules that memorize light irradiation events in the form of isomerization states at the subwavelength scale via near-field optics. In other words, we utilize the unique intrinsic attributes of photochromic materials [14] as well as the optical near-field interactions at the subwavelength scale [15,16] in order to overcome the diffraction limit of light.

Indeed, Nakagomi *et al.* succeeded in generating subwavelength-scale, autonomously formed complex structures in a photochromic material by localized near-field light excitation [17,18]. In their approach, the interplay between anisotropic deformation of photochromic molecules and localized near-field light yields a chain of transparent regions in the material, which we call a *memory pathway* hereafter. Furthermore, Uchiyama *et al.* generated Schubert polynomials, which are the foundation of combinatorial mathematics, based on the photon statistics observed through complex nanostructures formed in a photochromic material [19].

However, the unique capability of reconfigurable information storage in the form of photoisomerization at the subwavelength scale has not been fully utilized. That is, it is necessary to examine the addition and erasure of information on the nanometre scale, which we refer to as *history-dependent nano-photoisomerization* hereafter, which is also reflected in the title of this article. As mentioned above, history-dependence is of critical importance for solving complex problems, as observed in neuromorphic computing [4,5], multi-armed bandit problems [8], and reinforcement learning [20], among others.

Here we experimentally demonstrate sequential formation of multiple memory pathways in photochromic crystals by using a double-probe optical near-field microscope. In this microscope, one probe was located on the front side of a photochromic single crystal for local excitation, while the other was put on the rear side for observation. By carrying out localized near-field excitation at the front side, photoisomerization was introduced at the subwavelength scale, leading to the formation of a memory pathway, which was observed by scanning the optical fibre probe located on the rear side.

We examined history-dependent nano-photoisomerization by changing the spatial position of the front-side local photoexcitation. More specifically, local optical excitation was repeated twice, which we call the first and the second excitations, while the spatial position of the excitation was changed. As demonstrated later, a negative correlation was observed between optical near-field measurements on the surface of the photochromic material triggered by the first and the second excitations. That is to say, the near-field optical observation from the added memory pathway was mutually exclusive to that from the previously formed original memory pathway. We also confirmed that the first memory pathway was preserved after the addition of the second memory pathway. This result indicates that photoisomerization by an optical near-field in diarylethene crystals has history dependence, leading to a new paradigm of computations using light–matter interactions.

**Results and Discussion**

**Formation and observation of photoisomerization pathways.**

The photochromic molecule used in the present study was diarylethene, shown in Fig. 1a [21]. It has an open-ring isomer (**1o**) state and a closed-ring isomer (**1c**) state and can be made to undergo isomerization between these states by light irradiation. The open-ring isomer is transparent to visible light and is isomerized to the closed-ring isomer with UV light irradiation (Figs. 1a and 1b). The closed-ring isomer is opaque and blue-coloured and returns to the open-ring isomer by photoisomerization with visible light. The diarylethene molecule shows reversible photoisomerization even in the crystalline state, and both isomers are thermally stable [14].

We obtained a plate-like crystal by recrystallization using methanol as a solvent and coated the front and rear surfaces of the crystal with a platinum layer of 10 nm thickness by a sputter deposition method for observation using scanning tunnelling microscopy (STM). Figures 1c and 1d show the surface height profile of the prepared sample observed by the STM method. The surface consisted of a flat surface of molecular length order and a step structure of several nanometers. The surface structure was stable to photoisomerization, and the topography allowed a comparison of multiple SNOM images for positional identification. In addition, the typical scale of the surface structure was sufficiently smaller than the scales of the local excitation and local measurement, so the sample was suitable for the measurements in this study. See the *Methods* section for details of the crystal sample.

We measured optical input/output relations through the photochromic diarylethene single crystal using a double-probe scanning near-field optical microscope (SNOM) (Fig. 2a) [17]. One probe was an Au-coated tungsten probe for local excitation and was made to approach the front side surface of the crystal. The other was an optical fibre probe for local measurement of the optical near-field and was made to approach the rear side surface, as schematically shown in Fig. 2a. The optical fibre probe on the rear side was scanned for the two-dimensional measurement of the optical near-field to measure the input/output relations through the transparent nanostructure within the photochromic material, which is called the memory pathway. Figures 2b and 2c show scanning electron microscope (SEM) images of the Au-coated metal probe and the optical fibre probe, respectively. The relative transversal position difference of the tips of those two probes was adjusted within the measurement area (2 μm) by using our established method [17]. The relative misalignment of the two probes was estimated to be less than 1.5 μm. See the *Methods* section for details of the probe preparation and the alignment method.

The single crystal was initially coloured by irradiation with UV laser light (Vortran Laser Technology, Stradus 375-60: wavelength: 375 nm, power: ~100 mW/cm$^2$) for 30 mins. After moving the two probes 1 nm closer to the sample surface by utilizing the scanning tunnelling microscopy capability of the apparatus, we irradiated the sample surface with visible laser light (Melles Griot, 85-CGA-020: wavelength: 532 nm, power on the sample surface: several 100 μW/cm$^2$). With the excitation probe fixed, we measured an optical near-field map and a topography map on the rear-side surface of the sample at the same time. These maps were measured line by line in both forward and reverse scanning, making it possible to distinguish photon detection through the transparent nanostructures from background noise. Based on the acquired topography maps, positional deviations between consecutively measured SNOM data were calculated and used for positional correction when comparing the SNOM data. The measurement time for a pair of these images was about 6 hours. The scanning area was 2000 nm square, and the resolution was 256 x 256 pixels; hence the image resolution was 7.8 nm/pixel.

Here we discuss local optical excitation in this system and the formation of nano transparent pathways in the crystal [17,18]. The visible light incident on the sample surface was locally enhanced in the vicinity of the probe tip in a subwavelength region approximately the same size as the tip radius, which was approximately 100 nm (Fig. 2b). The enhanced local optical near-field excited and isomerized the photochromic molecules from the closed-ring isomer to the decoloured transparent open-ring isomer. An optical near-field has a high spatial frequency, corresponding to high momentum, so that molecules within the local excitation area are isomerized into the transparent state as a cluster from the initially coloured state. The generated transparent cluster has a spatial scale of several tens of nm [18,19] and works as an extended optical near-field source.

It should be noted that the local photoisomerization did not extend to the neighbouring area in an isotropic manner because the crystalline distortion induced by photoisomerization is anisotropic and acts on a spatial scale that is presumably comparable to the optical near field. The destination of the next photoisomerization branches out to multiple options to be selected. When one of the options is selected, the photoisomerization generates another mechanical distortion that constrains the subsequent photoisomerizations. In this manner, a chain of local photoisomerizations in the composite of the optical near-field and the mechanical distortion field spontaneously generates the transparent photoisomerized pathways, as schematically depicted in Fig. 2a. A SNOM image obtained by injecting photons from the same excitation point into the formed nano-photoisomerization pathways is shown in Fig. 2d. By applying an appropriate blur filter to the obtained optical near-field distribution map, input–output correlations at various scales within the pathways can be visualized (Fig. 2e).

It should be emphasized that the pathways were caused by the photoexcitation generated from a single photoexcitation point fixed on the front surface. And yet, complex subwavelength structures were observed via the optical near-field measurements from the rear surface. From this, we see that the optical output from an exit point via complex photoisomerization pathways in the nanostructure functionally corresponds to how an output is generated in a brain, which also contains complexity but inherits individuality or self. As mentioned already, we refer to the photoisomerization pathway as a *memory pathway*.

**History-dependent nano-photoisomerization as memory pathways.**

Figure 3a shows the experimental procedure used to investigate the history dependence in the sequential formation of multiple memory pathways. In Step (i), the first memory pathway was formed by local photoexcitation in the initially coloured crystal (Fig. 3a(i)). We call the first memory pathway **P1**. Also, the measured optical near-field map is referred to as **MP1** hereafter. In Step (ii), we shifted the local excitation point by 1 μm for the formation of the second memory pathway (**P2**) and measured the optical near-field map, which we call **MP2** (Fig. 3a(ii)). In Step (iii), we returned the local excitation point to the optical input point **P1**. We call the remeasured memory pathway **P1'** and the optical near-field map **MP1'** (Fig. 3a(iii)).

Figures 3b(i), (ii), and (iii) show the three optical near-field images obtained in the same area, **MP1**, **MP2**, **MP1'**, respectively. The distribution of the exit positions of the memory pathway was observed as subwavelength-scale bright spots in the optical near-field images. The number of photons detected per pixel was 22 on average in the measurements, with a maximum of 198 counts/pixel and a minimum of 3 counts/pixel. The dark count was 4.5 counts/pixel. To cancel out the noise and compensate for small changes in the experimental conditions, we normalized each line of the obtained image by the mean photon count and subsequently blurred the normalized image with a Gaussian filter of 3 pixels, corresponding to approximately 23 nm, which was the measurement resolution of the apparatus. Each image was 256 x 256 pixels for a 2-μm square region. Based on the topographic images taken simultaneously, we searched for the common area in the three images. See the *Methods* section for details.

In Fig. 3c, we extracted nanostructures of about 60 nm from the SNOM images by bandpass filtering for a spatial scale of 50–70 nm. Here we focused on a 250-nm-square region, indicated by the black square in each image in Fig. 3b, to investigate the relation between the three SNOM images concerning the nanostructures. The solid black circles depict the bright nanostructures observed in **MP1**, and the solid red circles indicate the bright nanostructures observed in **MP2**. In **MP1** (Fig. 3c(i)), the spots which were bright in **MP2** were dark, as indicated by the dashed red circles. On the contrary, in **MP2** (Fig. 3c(ii)), the spots which were bright in **MP1** were dark, as indicated by the dashed black circles. In **MP1'** (Fig 3c(iii)), the spots which were bright in **MP1** were also bright, and the spots which were bright in **MP2** were rather dark. A comparison of the three SNOM images indicated that the additional memory pathways induced by the second local excitation were formed while avoiding the memory pathways previously formed by the first local excitation.

We extracted structures with a spatial bandwidth of 10 nm at various scales (see the *Methods* section for details of the bandpass filter) and evaluated the correlation between the three images. The cross-correlations between [**MP1** vs **MP2**], [**MP1'** vs **MP2**], and [**MP1** vs **MP1'**] are shown in Figs. 4a, b, and c, respectively. The higher the correlation coefficient on the vertical axis, the more photons were observed via travelling through the same memory pathway, which evaluates the degree of coincidence of experienced multiple memory pathways. The horizontal axis represents the size of the nano-optical structures extracted from the SNOM images by the bandpass filter.

The comparison of **MP1** vs **MP2** shown in Fig. 4a showed negative correlations in the optical near-field regime, less than one-quarter of the wavelength, ~100 nm, which means that the nanometer-scale memory paths were different for different local optical excitation points. That is, the mechanical distortion field generated by the formation of the first memory pathway (**P1**) with photoisomerization exclusively constrained the formation of the second memory pathway (**P2**), leading to the formation of different paths. The comparison between **MP1'** and **MP2** shown in Fig. 4b also exhibited negative correlations, but their extents were relatively weak compared with Fig. 4a. This indicates that some of the photons from the excitation point **P1'** flowed into the newly formed **P2**. In other words, the memory of **P1** and the memory of **P2** generated a mixed memory in part.

Finally, we compared **MP1** and **MP1'**, which are related to local optical excitations given at the same point, but before and after adding another memory via **P2** (Fig. 4c). These structures were positively correlated; hence the structures were consistent for the same local optical excitation even after adding other memories by **P2**. The consistency indicates the conservation of the memory pathways.

Here we make some additional remarks about the experiments. After finishing the first memory examination at **P1**, we irradiated the sample with UV light for 90 min with the intention of completely resetting the material before starting the second memory examination at **P2**. Furthermore, we added local photoexcitations at other points and carried out the SNOM measurements four times in addition to the three SNOM measurements mentioned above (**MP1**, **MP2**, and **MP1'**). The first remark is that we could not find apparent nanostructures in the optical near-field maps regarding these other photoexcitations. We speculated that the local excitation point was not fixed well enough during the optical near-field measurements; hence, the memory pathways were not formed well enough to be observed on the rear side. In the case of the three photoexcitations (**P1**, **P2**, **P1'**), we consider that the local excitation probe was well fixed at the same point so that the pathways from the front side to the rear side were sufficiently formed with photoisomerization and were observed as optical near-fields on the rear side.

The second remark is about the observation that **MP1** and **MP1'** exhibited a positive correlation even though a 90 min recolouring process with UV irradiation was conducted between the observation of **MP1** and **MP1'**. That is, the original memory pathway was not completely erased. This indicates that the memory pathway formed on the nanometer-scale by local excitation was resistant to or robust against the far-field light irradiation. Such resistance may be supported by the mechanical distortion field around the memory pathway. In future research, we will investigate whether the erasing begins from fine structures or whether the memory is retained as peripheral distortion even after the isomerization path disappears.

The third remark is about the relation to the brain and neuromorphic computing [4,5]. As mentioned in the previous section, the photoisomerization memory pathways are formed in the balance between the optical near-field and the mechanical-distortion field. The addition of a new memory was conducted under the mechanical distortion field around the existing memory (Fig. 5a). The new memory pathways tend to pass through gaps in existing structures so that the photoexcitation chain along photoisomerized regions remains independent to avoid the mixing of the optical near-field interaction networks. Such a complex and spontaneous memorization with history dependence is like a memory function formed by adjusting the synaptic connections in the brain (Fig. 5b). In the brain, a memory affects the formation of another memory in a neural network with accompanying environments of the chemical field by glial cells [22]. In nano-photochromism in photochromic crystals, a mechanical distortion field generated by a memory produces a probability distribution of photoisomerization as the environments for the subsequent memories on the nanometer scale.

The final remark regards mathematical insights into such history-dependent nano-isomerization. As mentioned in the Introduction, Uchiyama *et al.* generated a series of order structures as the matrices deduced from the memory pathways obtained for a fixed single excitation point [19]. The generated series exhibited diversity and looked almost random; however, they were all correlated because the near-field photon statistics were obtained by a fixed-position local excitation [23]. The double memory pathways observed in the present study provide the order structures with *two* order structures being correlated. Detailed analysis of this will be an interesting topic for future work. Furthermore, we can explore interesting applications, such as adversarial relationships, which have been intensively studied in recent research on artificial intelligence [24,25].

## Conclusion

We demonstrated the addition of memory pathways and examined the relation of the memories through the optical input/output relations observed by a double-probe scanning near-field optical microscope to investigate the history dependence in the optical-near-field-induced formation of photoisomerization memory pathways in a photochromic single crystal. An additional memory pathway was formed exclusively with respect to an originally formed memory pathway. The first memory did not disappear but mixed with the new memory partly. Such a history-dependent memory is formed spontaneously in the local balance between the optical near-field and mechanical distortion field, which is similar to a memory formed by tuning the synaptic connections in the brain. These results suggest the usefulness of diarylethene crystals as a dynamic memory structure in non-von-Neumann-type devices and systems. In future work, we will further study dynamic and highly multiplexed, brain-like memory architectures for intelligent functionalities by exploiting the interplay between photochromism and near-field optics.

## Acknowledgements

This work was supported in part by the CREST project (JPMJCR17N2) funded by the Japan Science and Technology Agency and Grants-in-Aid for Scientific Research (JP20H00233, JP21K04925) funded by the Japan Society for the Promotion of Science.

**Methods**

**Sample preparation**

The molecule used in this study was a diarylethene, 1,2-bis(2,4-dimethyl-5-phenyl-3-thienyl)perfluorocyclopentene, whose molecular formula is shown in Fig. 1a. In the open-ring isomer **1o**, the molecule is transparent to visible light and has a long side of 1.41 nm. In the closed-ring isomer **1c**, the crystals of **1c** are opaque and blue, with a long side of 1.39 nm. The absorption spectra of these two isomers are shown in Fig. 1b. Upon irradiation with UV light, the open-ring isomer **1o** isomerizes to the closed ring isomer **1c**. The closed-ring isomer **1c** is isomerized to the open-ring isomer **1o** upon irradiation with visible light. The opening and closing of the ring structure in the centre of the molecule changes the conjugation length of the molecule, which in turn changes the absorption spectrum. The symmetry of the crystal structure is the same for the coloured and decoloured states so that photoisomerization proceeds without breaking the crystal lattice.

In this experiment, we used a plate-like single crystal produced by recrystallization from a hexane solution. The thickness was about 0.17 mm, and the size of the largest planar surface ((100)-surface) was about 0.5 mm. For scanning tunnelling microscopy (STM) control, the surface was made conductive by coating it with a thin Pt layer about 10 nm thick.

**Probe preparation and alignment method**

The tip of the metal probe was sharpened by an electrochemical method down to 100 nm with an NaOH solution and coated with a 10 nm-thick gold layer by a sputtering method. The optical fibre tip was sharpened down to less than 10 nm by chemical etching with buffered hydrofluoric acid. The tip was coated with a 10 nm-thick platinum layer for position control by STM, and the other part was coated with a 50 nm-thick platinum layer for preventing the detection of propagating light. The final radius of curvature of the tip was about 10 nm, which determined the spatial resolution of the optical near-field measurements.

The method for aligning the two probe tips was reported in a previous paper [17]. In short, we confirmed the alignment of the two tips through the capacitance between the two probes in two dimensions. After the alignment, we brought the two probe tips close together and confirmed their relative positions with the resolution of the STM method.

**Determination of common area in SNOM images**

We measured the optical near-field map and the topographical map at the same place and the same time. In scanning probe measurement at room temperature, thermal drift of the relative position of the sample and the tip is inevitable. In this experiment, we confirmed that the drift was less than the size of the local photoexcitation (~100 nm) during the measurements shown in Fig. 3 based on the STM images. To subtract the drift in the interval of the three measurements, we found the common area in the STM images to compare the SNOM images taken in the same area. The resolution for the subtraction was determined by that of the STM image, namely, 7.8 nm.

**Extraction of a certain spatial structure from the SNOM images**

The SNOM images had information about the optical near-field intensity in the vicinity of the scanning probe and also the propagating light distribution that could not be removed. To remove the modulation by the incident light (temporal changes in detector sensitivity and laser intensity), we subtracted the mean intensity from each line of the SNOM images. In the SNOM images modified by the subtraction (shown in Fig. 3b), large structures with a scale equal to the propagating light wavelength, about 500 nm, were observed. To investigate the correlation between the SNOM images, we should separate the optical near-field components from the propagating light components so that we can extract the appropriate scale information from the SNOM images by a bandpass filter. In detail, when the obtained SNOM image is represented by a two-dimensional matrix $M$, we denote the Gaussian filtered matrix with a standard deviation of $L$ (nm) by $M_{G(L)}$. To extract the structure from $L_1$ (nm) to $L_2$ (nm), $MB(L_1, L_2)$, we calculate as follows:

$$MB(L_1, L_2) = \left(M - M_{G(L_2)}\right)_{G(L_1)},$$

assuming $L_1 < L_2$. Fig. 3c shows $MB(50 \text{ nm}, 70 \text{ nm})$ with a bandwidth of 20 nm. In Fig. 4, the point for 100 nm (two times of the centre standard deviation of 50 nm), for example, was calculated by $MB(45 \text{ nm}, 55 \text{ nm})$ with a bandwidth of 10 nm.

**Data availability**

The data that support the findings of this study are available from the corresponding author upon reasonable request.

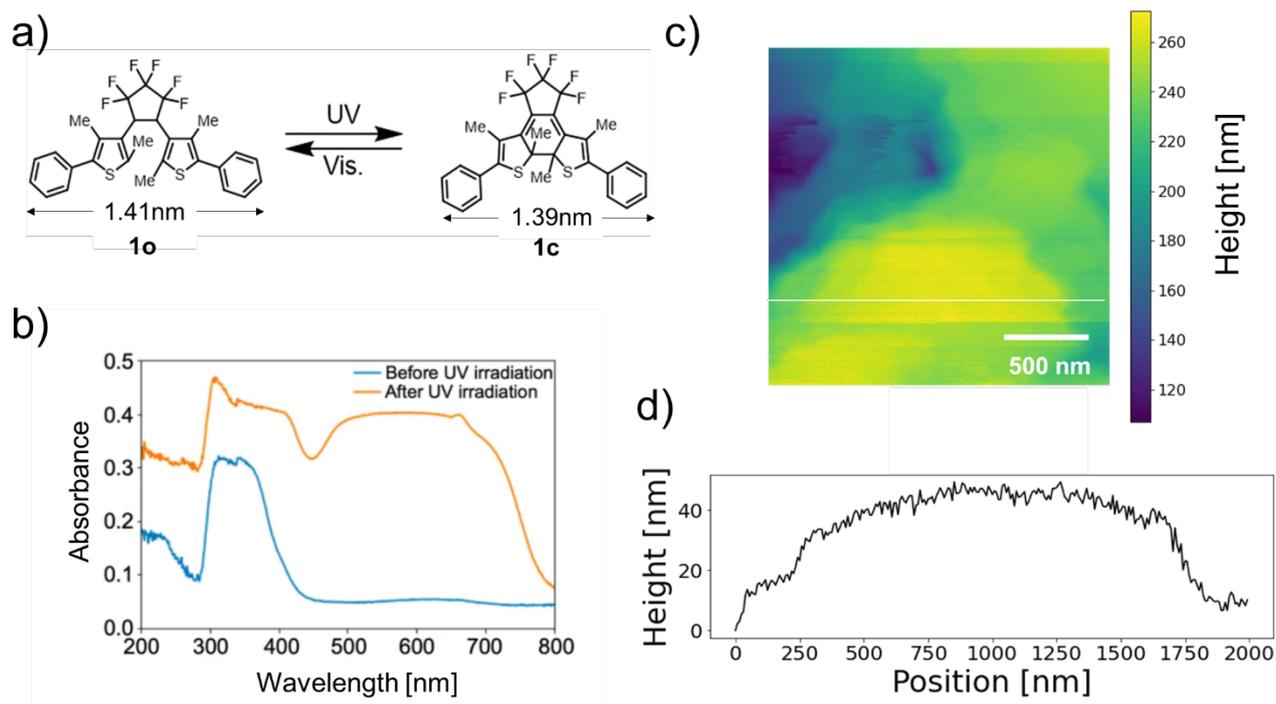

**Fig. 1 | Optical properties and surface structure of photoisomeric diarylethene crystal. a,** Photochromic diarylethene molecules and the photoisomerizations. **b,** The absorbance spectra for the transparent and coloured photochromic crystals. **c,** STM topography on the sample surface. **d,** The height profile along the white line depicted in **c**.

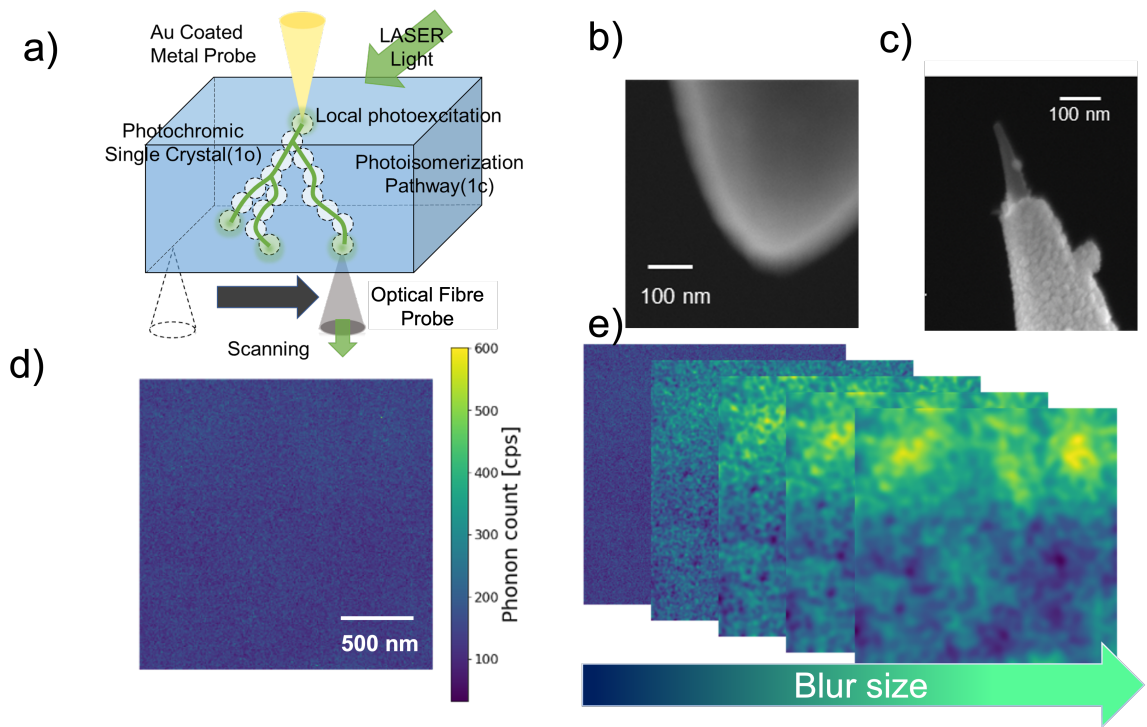

**Fig. 2 | Local excitation and local photon detection on a photochromic single crystal. a,** Schematic diagram of the experiment. Local excitation by metal probe generates the photoisomerization pathways in the crystal, and the photons travelling through the pathway are detected by the optical fibre probe on the rear side. **b,** Metal probe for local photoexcitation used in the experiment. **c,** Optical fibre probe for local measurement of the optical near-field. **d,** one of the obtained SNOM images as output of the photoisomerized pathways formed with the chain of anisotropic photoisomerizations. e, Visualization of the local optical structure with a blur filter of the corresponding scale.

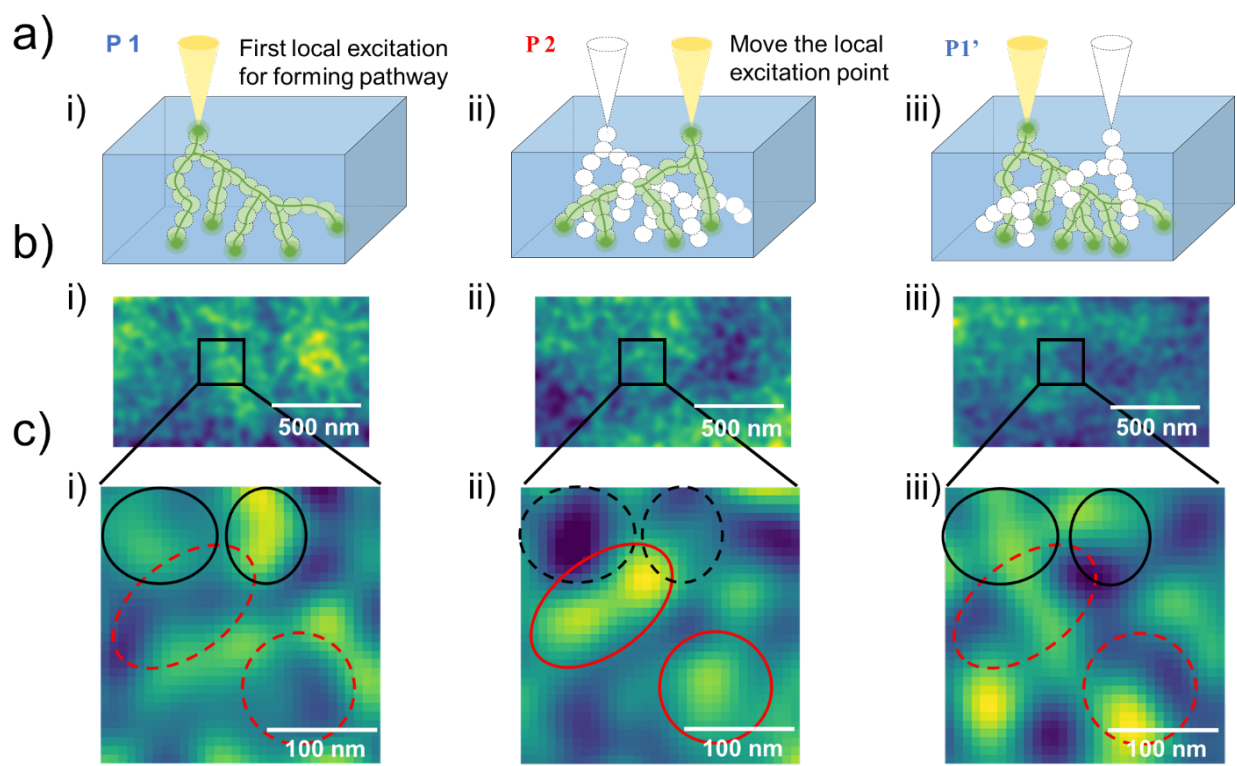

**Fig. 3 | Measurements of multiple memories. a,** Measurement procedure of history dependence. **b,** SNOM images for the three measurements in the same area. **c,** Magnified SNOM images of the 250 nm square area indicated by the black boxes in b.

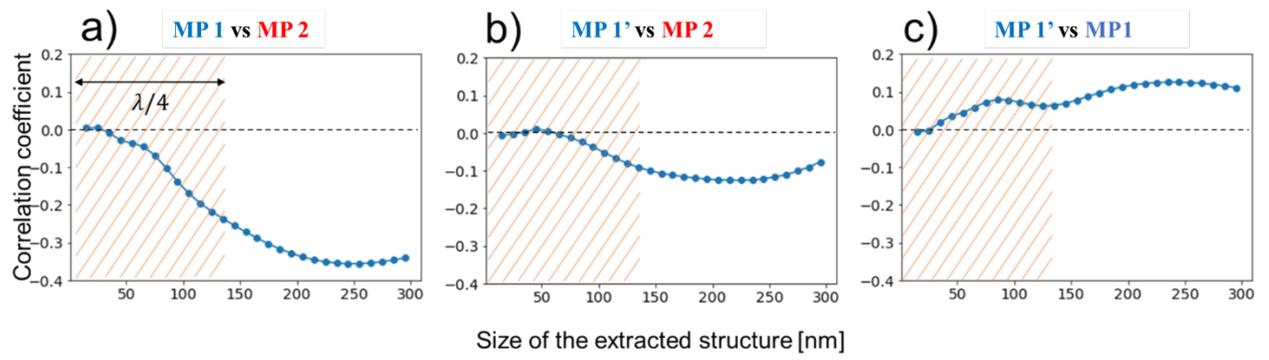

**Fig. 4 | Cross-correlations between the three memory pathways. a,** Cross-correlation of **MP1** and **MP2**. **b,** Cross-correlation of **MP1'** and **MP2**. **c,** Cross-correlation of **MP1** and **MP1'**.

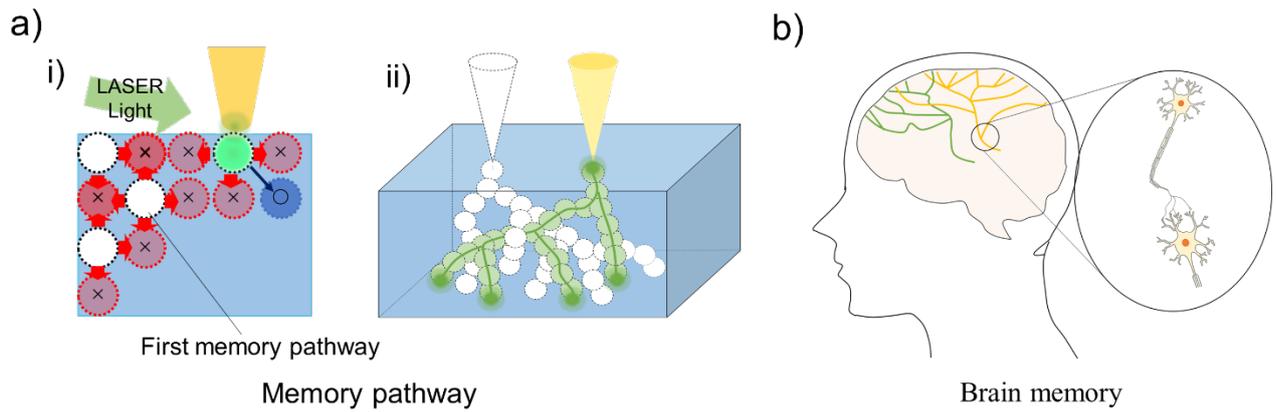

**Fig. 5 Model of multiple memories in photochromic crystal. a,** Two memory pathways formed with history effects. (i) The white circles mean the first memory. The second memory is formed under the local excitation from the metal tip to avoid the regions forbidden from being photoisomerized. (ii) The expected relations of the two memories. **b,** Analogy to the two memory circuits formed by synaptic connections triggered by two events in the brain.